# SUSY Via Magnetic Monopoles and Chirality

David R Winn
Fairfield University Physics
Fairfield, CT 06824-5195 USA
winn@fairfield.edu

*Abstract:*
A SUSY is proposed that interchanges bosons-fermions, and also transforms particle electric charge to magnetic charge, chromoelectric charge to chromomagnetic charge, and interchanges left-right handed couplings, leaving the masses between partners intact, and without R-parity. Constraints and consequences are discussed.

*Introduction:*
We have compelling evidence for Dark Energy and Dark Matter, but the particle or field nature of DE and DM is still only conjecture. The preponderance of DE and DM over ordinary matter, and of matter over anti-matter, remains not fully explained. Magnetic Monopoles of magnetic charge $g=n (e/2\alpha)$ have not been found, despite compelling reasons for them to exist (Dirac charge quantization[1]; t'Hooft-Polyakov and topological arguments[2,3]), although non-perturbative QED/QMD may prevent unbound monopoles from being pair-produced, and so very difficult to find in isolation, much like isolated quarks. Considering present LHC experimental limits, softly-broken SUSY with R-parity, if it exists, seems at best to be at a very high mass scale, and so the stability of the Higgs mass remains a puzzle. Therefore, we propose a SUSY with unbroken masses, but in addition to interchanging bosons-fermions, it changes electric charge to the pseudoscalar magnetic charge, chromo-electric charge to chromo-magnetic charge, and the handedness of the SUSY partners L-R is reversed.

An Ansatz - magneto-chiral-SUSY:
- Transforms vectors into spinors, and spinors into scalars, as usual.
- Transforms electric charge e into magnetic charge $g = n (e/2\alpha)$ – there are no electrically charged superpartners, only neutral or magnetic charged partners. The magnetic charges of the squarks are multiples of $3g$ [4].
- The chromo fields of the quarks and gluons are also replaced by chromomagnetic fields for the squarks and gluinos.
- The supersymmetric masses are degenerate with the normal masses.
- The handedness of the couplings in the SUSY sector, L vs R, is reversed.
- R-parity conservation is not needed.

*Issues & Consequences of the mcSUSY ansatz:*

*1) Higgs Mass Stability:*
To first order, the cancellation of divergences in the Higgs mass becomes exact, as the superpartners have the same masses.

*2) Magnetic SUSY Partners are mainly Hidden or Separated from ordinary matter:*

The magnetically charged and chromo-magnetically charged SUSY partners cannot transform into their partners and vice-versa in first-order couplings, even if one is off-mass-shell, since at the interaction vertices both electric AND magnetic charges must be separately conserved, intrinsic angular momentum must be conserved, and the handedness of the interactions is changed at the interaction vertices, and mass/energy and momentum are conserved in the final states. An electrically charged particle and its magnetic superpartner cannot be at the same vertex unless there are cancelling charges; i.e. separately, the net electric charge AND magnetic charge into and out of an interaction must not change. For example, a magnetic $\tilde{e}$ cannot convert to e+$\tilde{\gamma}$ alone. Similarly, chromomagnetic gluinos are not produced in single diagrams from quarks or gluons. The result of magnetic and chromomagnetic charge, the same masses, and the inverted handedness of the couplings is that the SUSY sector is well-hidden.

*3) Proton Decay:* The proton lifetime is not diminished by magneto-SUSY mediated decays, and without requiring R-parity for this trick, since traditional SUSY diagrams (such as those for p → K+ν) are excluded by conservation of both charges in the interaction vertices.

*4) Magnetic Monopoles:*
So where is the 0.511 MeV magnetic $\tilde{e}$ ? It is widely believed that magnetic monopoles (i.e. here, the "charged" SUSY partners) cannot be pair-produced into free particles in analogous ways to pairs of electrically charged particles, since, much like quarks & gluons, the magnetic forces for magnetic charges of ng~ n(e/2α) are non-perturbative and in effect confining[5]. Because of the large size of the magnetic charge, this is a strong coupling problem for which perturbation theory cannot be trusted; obtaining quantitative estimates of the production cross-sections of monopoles, free or bound, remains an open one. The resulting strong magnetic force between any pair-produced monopoles explains the absence of detected monopoles heretofore - since they pair up to g=0 and annihilate back to our matter when they are produced, and/or quickly screen themselves by creating opposite charged pairs, similarly to quark hadronization. The "ionization" energy between pairs of monopoles created over the space of a nuclear radii in a p-p or p-nucleus collision is naively >1TeV - but this is a low estimate because of uncontrollably large non-perturbative effects. As an example, the naïve "Rydberg" energy of a bound selectron-muino system exceeds 1-2 GeV, even if they have the minimal magnetic charge ±g, and the resulting spinless g=0 "atom" becomes essentially massless. The size of the magnetic force explains much of the absence of this amusing version of SUSY in experiments, and indeed also why even in the cosmos bare monopoles are exceedingly rare, if at all. We will discuss g=0 monopole bound states that may be created in ordinary matter interactions below. We note that the monopole particle spectrum of SUSY monopoles proposed here seems more natural than if a single type of monopole was the only magnetic charge created in nature.

*5) QED Effects in Known Phenomena:*
Higher order QED phenomena with normal matter are affected at very low levels by virtual (mostly scalar) magnetic charges. For example the Lamb shift would not be affected by magnetic (pseudo)scalar right-handed selectron bubbles, as the energy shift is due to an electric field shift, even for a selectron of normal electron mass[6]. Vacuum polarization phenomena with spinor magnetic monopoles affecting ordinary electronic matter are highly suppressed or vanish. Caution to these initial conclusions is necessary since the magnetic selectron is a (complex? Pseudo?) scalar. The Muon g-2 anomaly is controversial: an early calculation showed that a spinor monopole mass must have M> 60 GeV(corrected from the original 120 GeV), to be consistent with the g-2 magnetic moment anomaly as measured by experiment[7]. However, this result is highly criticized due to the non-perturbative nature of monopole bubbles and higher diagrams, and that the lower limit is either 2 GeV or vanishes[8]. We note calculations of monopole effects use the existence of only a single type of spinor monopole with the lowest magnetic charge. The lowest mass and only spinor monopole is the right-handed $\widetilde{W}$, and the lowest mass scalar or pseudoscalar monopole is the right-handed $\tilde{e}$. The other monopoles are all scalars or pseudo scalars, the heaviest being the $\tilde{t}$ (unless there are non-minimal charged H, and therefore the $\widetilde{H}$ superpartners are magnetically charged). In any case, a more thorough investigation of existing QED phenomena, including the effects of scalar and pseudoscalar right-handed $\tilde{l}$ and $\tilde{q}$ with normal masses, an $\widetilde{H}$ mass of ~125 GeV, a spinor magnetic RH $\widetilde{W}$, a massless spinor $\tilde{\gamma}$, a neutral spinor $\tilde{Z}$ of opposite handedness, is required to determine if present experimental measurements are in agreement.

*6) Invisible Z-width:*
Left-handed scalar $\tilde{\nu}$'s should contribute about ½ of the invisible width, about 9% of the width of the Z. This is experimentally ruled out by ~2 orders of magnitude. If we demand instead that L-R handedness is reversed in the SUSY sector, then this vanishes for a fully right-handed $\tilde{\nu}$ [9],[10]. A more general schematic model of the $\tilde{\nu}$'s is a superposition of left and right-handed states: $\tilde{\nu} = - \tilde{\nu}_L \sin\theta + \tilde{\nu}_R \cos\theta$. The $\tilde{\nu}$ contribution to the Z width $\delta\Gamma$ is proportional to $\sin^4\theta$ x ($\Gamma\nu$ = 167 MeV, the Z width to ordinary $\nu$'s). If $\sin\theta$ <0.4, the mixing with LH couplings is allowed. Light, sterile mixed but mainly RH $\nu$'s and $\tilde{\nu}$ 's are plausible candidates for dark matter [11]. An alternative viewpoint: scalar $\tilde{\nu}$ 's are unusual particles. Presumably the $\tilde{\nu}$ complex scalar wavefunction can be a superposition of neutrino and antineutrino (i.e. they have the same masses) spinor wavefunctions, and it is possible that the square of the amplitude vanishes. In effect, they have already contributed to the Z invisible width, half $\nu$ & half $\tilde{\nu}$. We note/acknowledge that the level of mixing between L,R handedness may be important to cosmic abundances of $\tilde{\nu}$ dark matter; mainly RH scalar $\tilde{\nu}$ 's may form Bose condensates – hence Dark Matter - in the cosmos. (We note in passing that the Higgs and the photon could be similarly thought of as a complex linear combinations of $\widetilde{H}$ or $\tilde{\gamma}$ – see below).

*7) High Energy Limits on Monopole Masses:*
It has been proposed that virtual monopoles can mediate high energy processes which give rise to multi-real-photon final-states. The lower limits on spinor monopole masses estimated from experimental measurements of Z→ 3$\gamma$ [12] (~400 GeV monopole mass lower limit) or $\gamma$–$\gamma$ scattering[13] (production of two real photons with high transverse momenta by the collision of two photons produced either from e+e− or q-q collisions) experiments [14],[15] that indicate lower limits on fermion monopole masses 500-600 GeV, have been strongly disputed[16]. These calculations of monopole-mediated processes which result in real photons are based on overly simple applications of electromagnetic duality;

the resulting cross sections cannot be valid, because unitarity is violated for monopole masses lower than quoted limits; the processes are subject to enormous, uncontrollable radiative corrections, and indeed, preliminary analysis shows that amplitudes with real photons created by spinor monopole square loops vanishes. Whether this holds for R-handed (pseudo)scalar monopoles is not yet known.

*8) Rates of Monopole Couplings to Photons and Z's:*
An important question is whether the rate of e+e- into leptons and hadrons, or the width of the Z into charged particles, is changed by the existence of the proposed spectrum of SUSY monopoles. Since magnetically charged SUSY particles re-annihilate before they can be free, then likely not, and, moreover, the handedness is reversed, as in the discussion of the invisible Z width above. However, it is possible that a proper analysis of experimental limits rules this model out.

*9) Controversial or New Interactions/Diagrams:*
(i) New 4-point scalar coupling/interactions, of, for example, the 4 legs $\tilde{e}^+\tilde{e}^-\,\tilde{\nu}\tilde{\nu}$, analogous to the 4-point couplings of the Higgs - Are they allowed?
(ii) Dimension-4 couplings of magnetic+antimagnetic charged scalars (say selectron+spositron, or $\tilde{q}$ –anti $\tilde{q}$ ) to $\tilde{\gamma}$-anti$\tilde{\gamma}$ spinors. This might be a "blob" of non-perturbative magnetic couplings with 2 scalar and 2 spinor legs.
(iii) Other Yukawa or interaction terms – unknown/unproposed at present.

*10) Photinos:*
The massless spinor $\tilde{\gamma}$ is profoundly indifferent to others. However, we emphasize that the photon can be viewed as a complex linear combination of massless $\tilde{\gamma}$, and this may have further consequences, if unusual higher dimension couplings are allowed. The massless $\tilde{\gamma}$, if present in the universe at large with the same or larger density as photons, but 2 spinor types, forms a fully relativistic Fermi Gas with a negative pressure similar to that indicated by observational Dark Energy(see section 20 below). Photinos may be made by $\gamma\gamma \to \tilde{\gamma}\tilde{\gamma}$ scattering, and by photons interacting with black holes. Photinos with frequencies approaching zero may fill the universe.

*11) Higgs Production and Decay:*
The H→γ-γ rate may be larger, since the mass of the scalar $\tilde{t}$ is the same as the t and the spinor $\tilde{W}$ may also contribute to the triangle diagram of the decay. Effects of magnetic chiral unbroken mass SUSY on the other production mechanisms and BR of the Higgs need to be analyzed; the results of the analysis may rule out this proposal.

*12) Sleptons:*
The lepton sector can communicate with sleptons, since there are neutral s/leptons= $\tilde{\nu}/\nu$'s, and charge=g leptons, so that $\tilde{W}$ (charge ±g) can be exchanged (this is unlike the squark sector). The SUSY sleptons can be converted into each other with interactions that look like weak decays with spinor $\tilde{W}'s$; to first order they are immune to the W by electro-magneto charge conservation. For example, a scalar magnetic $\tilde{\mu}$ of charge g can decay to a spinor magnetic $\tilde{W}$ charge g, and a normal $\nu_\mu$. So, for example, a free or bound $\tilde{\mu}$ can decay into a $\tilde{e} + \nu_e + \nu_\mu$. Another example: $\nu+\bar{\nu} \to \tilde{l} + \bar{\tilde{l}}$. These processes are in principle observable. Slepton pair production at colliders or in cosmic rays may produce bound states: if a pair of $\tilde{\mu}$-$\tilde{\mu}$ is made at LHC, a pair of $\tilde{e}'s$ could materialize to make them g=0 $\tilde{\mu}$-$\tilde{e}$ bound states, similarly to a pair of produced quarks "dressing" themselves - a scalar $\tilde{\mu}$ + $\tilde{e}$ bound state naively would have a binding energy ~1.5-2 GeV, "larger" than its mass - i.e. this is a sine qua non of entering a non-perturbative regime. Such a composite is allowed to decay into a $\tilde{\nu}$ -$\tilde{\nu}$ or ν–ν pair – *assuming that s/neutrinos are lighter than the bound state.* Similarly, even a $\tilde{\tau}$ − $\tilde{\mu}$ bound state would likely be near zero mass, and would look like missing energy at colliders. It is amusing to consider the electric dipole moment of a (quasi) stable bound state of monopoles - analogue of the magnetic moments of hadrons (or atoms), with the caveat that everything is boson-like, except when a $\widetilde{W}$ is involved. Perhaps this should be searched for with highly inhomogeneous electric fields - like the tips of lightning rods. The analog of the Bohr Magneton – the "Bohr Electroton", of, say, the $\tilde{e}$ is ~0, since it is a (pseudo)scalar. A consideration that may make this detectable or ruled out: what is not really calculable is magnetic excited states of slepton-antislepton bound states (and squarks also) - and whether the radiation from de-excitation is observable or already ruled out, either here on earth or in the cosmos.

*13) Sneutrinos:*
These handed complex scalars may form Bose condensates as a part of Dark Matter. This is unlike most SUSY models, where the masses are large. Are they Majorana-like or Dirac-like? As discussed above and in the references, if the mass eigenstates have only or mainly right-handed couplings, they survive first order annihilation in the cosmos. An interesting question is distinguishing mixing between themselves and mixings of ordinary neutrinos and antineutrinos, and with ordinary neutrinos, since they have the same masses. If superpositions of neutrinos are indistinguishable from $\tilde{\nu}'s$ they will affect neutrino oscillations, and $\tilde{\nu}'s$ may be an effective sterile 3+2/3 component that solves some contradictory oscillation data[17]. Depending upon phases, $\tilde{\nu}'s$ are a promising avenue for lepton-violating processes and CP violation.

*14) Squarks and Gluinos:*
Scalar magnetic & chromomagnetic $\tilde{q}'s$ and spinor chromomagnetic $\tilde{g}$ are clearly controversial. Since all the scalar $\tilde{q}$ are magnetically and chromomagnetically charged, first order interaction to normal matter is through the neutrals: γ, Z, Zino, H or the $\widetilde{H}$, or by combinations of g and $\tilde{g}$. The electrically charged W and magnetically charged spinor $\widetilde{W}$ have difficulty coupling to $\tilde{q}$ or charged $\tilde{l}$, or quarks and charged leptons, respectively. In the Dirac formulation, the smaller the electric charge, the larger the magnetic charge. If we follow Dirac's argument for quarks transformed to magnetic $\tilde{q}$, the $\tilde{q}$ magnetic charges are -3g and +3/2 g. So for example, naïvely, a -1/3 d-quark would have a magnetic charged $\tilde{d}$ =-3g squark (g~137/2 e), and a +2/3 u would have a magnetic charge superpartner $\tilde{u}$ =+3/2g. However, following arguments delineated by Preskill[18], since the $\tilde{q}'s$ presumably are confined, the magnetic charges must be integer multiples of g. As examples: one possibility: (i) multiply by 2: $\tilde{d}$= -6g and $\tilde{u}$ =+3 g, or (ii) choose the least integral charge for the $\tilde{d}$ = -1g, and for the $\tilde{u}$ = +2g, to preserve the 2:1 charge ratio. So imagine a 1st order weak decay $\tilde{u}$ to $\tilde{d}$. The magnetic charge difference changing one to the other is either 9g or 3g. So this decay by weak processes is first order forbidden; i.e.

there is no analog of the W that has such a large charge except for bound states/effective particles. That weak decays seem forbidden except by flavor-changing neutral Higgs exchange in the magneto-susy sector may lead to unusual stable (Dark?) matter. We note that flavor changing neutral Higgs mediated decays are highly suppressed in our world by the data from kaon and muon decay, and double-beta decay. The consequences of chromomagnetic $\tilde{g}$, spinors, are not clear - for now we will assume they can bind $\tilde{q}$'s, and possibly q's. What is missing? A 3-spinor $\tilde{g}$ vertex, unlike gluons. A default position is to simply assume the scalar chromomagnetic $\tilde{q}$ can form bound states like regular quarks, with a glue of $\tilde{g}$'s and/or g's. What seems unknown – can chromomagnetic $\tilde{q}$'s, $\tilde{g}$'s bind also with quarks/gluons, and as such is this already ruled out by the known hadron spectra or monopole searches?

*15) Dark Matter as Squark/Gluino or Slepton Stable Matter?*
Where are the Shadrons? Amusing consequences of magnetic chiral SUSY: perhaps there is a scalar sproton made of the squarks [$\tilde{u}\tilde{u}$d], with either case (i) as above, which gives a sproton charge g=0, or, case (ii) as above, g=+3. The neutronino [$\tilde{u}\tilde{d}\tilde{d}$] squarks in case (i) is g = -9 or in case (ii) g=0. As an example, select Case (ii): a scalar (assuming the spinor $\tilde{g}$'s do not contribute to the spin) neutronino, with q=0, g=0 and spin=0(?), with a (pseudo?) (scalar?) sproton with magnetic charge ~ +205e or greater. If the latter exists, it would make bosonic hydrogen-lithium, a shydrogen with 3 scalar selectrons in the same ground state orbit, naively with a ground state radius ~nucleon radius, and binding energies of many GeV. Whether a charge 3g sdeuteron could exist, or which is lighter: the neutronino or the sproton (the magnetic energy of a charged sbaryon could be huge) are interesting, especially if the SUSY transform does transform color charge as well as electric charge. The 3-$\tilde{g}$ coupling (and other gluon couplings), for example, likely does not exist for spinor chromomagnetic $\tilde{g}$. It is unclear at this time what the consequences of chromomagnetic spinor gluinos would be on the spin of the analogs to hadrons - the sbaryons/barynos, or the smesons/mesinos. The charged Spion/Skaon could be stable to first order, but charged +/-3g; pair production would result in purely smesonic atoms that would annihilate back to ordinary matter. The magnetically neutral skaons, $\widetilde{bs}$ or other neutral smesons might also at least quasi-stable, since again there is no analog of the W that flavor-changes – the $\widetilde{W}$ is charged g=+/-1 and is a spinor, and so $\tilde{q}$ with charges differing by more than ±g do not happen. In principle, a flavor-changing H or $\tilde{H}$ pair exchange between $\tilde{q}$'s would allow the heavier smesons and sbaryons to decay – albeit highly suppressed in the SM - so maybe there are quasi-long lived g=0 shadrons. The production of neutral sbaryons or smesons seems possible. If there is shadronic matter, the experimental consequences needs sorting out:
- Neutral shadrons made at accelerators (or by the deitron) could look like normal long-lived neutral hadrons (K-longs, neutrons, lambda…) in a typical HEP or cosmic ray experiment, and be hard to distinguish; calorimeter energy would result. However, the cross-sections of chromomagnetic-matter with chromo-matter are not discussed here, and are likely unknown, and could be small.
- The electric dipole moment of magnetic sprotons, sneutrons, or other sbaryons/barynos - could be large, and possible to search for, depending upon whether spinor $g$ dominate the spin, or the scalar $\tilde{q}$.

- Purely sleptonic satoms, such as $\tilde{e} + \tilde{\tau}$, $\tilde{\mu} + \tilde{\tau}$, $\tilde{\mu} + \tilde{e}$ bound states, may have masses after binding less than s/neutrinos masses, and so may be stable or quasi-stable. It is amusing to imagine these bound scalar states as degenerate in mass with the neutrino or sneutrino mass eigenstates.
- $\tilde{\mu}$ or $\tilde{\tau}$ replacing $\tilde{e}$ in satoms may also exist, where the magnetic bound energy disallows decays.
- It is unclear what the heaviest stable magnetic dark sbaryon ($\widetilde{ttt}$?) or heaviest stable charged smeson ($\widetilde{bt}$?) and on downward would be, if any (i.e. whether chromomagnetism allows them to exist?) – and would depend on whether flavor-changing H or $\tilde{H}$ or higher order diagrams enable decays, but, together with possible stable $\tilde{l}$ magnetic bound states, or possible neutral snucleon- or smeson-like states, may constitute a complicated spectrum of dark matter candidates.
- Magnetically neutral Satoms in the ground state would likely be extraordinarily difficult to detect, the coupling to normal nuclei as yet unknown, and masses less than ~1GeV, and crudely speaking the deeply bound selectrons would be orbiting essentially at the surface of the sproton/snucleus.
- Induced electric dipole moments of smatter would be small, $\mathbf{p}_{shydrogen} \sim gh/2\pi m$.
- If excited, satoms could emit gamma-rays possibly up to 100's GeV. Shydrogen, if a charge 3g sproton and 3 g = -1 $\tilde{e}$'s would have bosonic rather than fermionic energy levels. This may have bearing on astrophysical observations.
- The existence or properties of snuclei are unknown at this point, with the bizarre (or likely boring) chemistry of the resulting satoms, where all the $\tilde{e}$'s are in the same ground orbit.
- Because of the relative stability of the SUSY sector, it is possible to imagine that this is why the amount of Dark Matter exceeds ordinary matter by an order of magnitude, if made freely in the primordial soup.
- If Smatter are Dark Matter candidates, they likely explain the paucity of signals in underground experiments, heretofore. The interactions of these relatively low mass neutral composites with atoms, atomic electrons or nuclei can be made very small indeed, but depend especially on the actual properties of magnetic/chromomagnetic $\tilde{q}$-$\tilde{g}$ snuclei.
- RH (or quasi-sterile mixtures) of $\tilde{\nu}/\nu$ sector condensates also contribute to Dark Matter, and may provide the bulk of it, with almost nil register in searches except for mixed states or oscillations.
- The existence of massless $\tilde{\gamma}$'s or partner magnetic smatter may affect stellar structure or evolution/supernovae, primordial H/D/He/Li abundances, radiation from black hole infalls, and other energetic processes in cosmology.

*16) Photon/Photino, Z/Zino, Higgs/Higgsino, and Axion/Axinos-Saxinos:* the photon, the Z, the Higgs and the Axion are degenerate with linear combinations of their same-mass SUSY partners in this model. The possibly rich ramifications of this mixing should be explored.

*17) Wino:* The $\widetilde{W}$ of charge g cannot participate in $\tilde{q}$ transformations with $\Delta g > \pm g$, and is the only magnetically charged spinor. It decays mainly into $\tilde{l} + \bar{\nu}$. A normal Z* may couple to a pair of $\widetilde{W}$. The $\widetilde{W}$ electric dipole moment ("Bohr Electroton") is $p_E \sim eh/8\pi\alpha cm_W \sim 10^{-33}$ J/(V/m).

*18) Composite Dyons:* The composite of an integer-spin monopole and an integer-spin charged particle can be a Dyon with half-odd-integer spin[19]. Do they exist as bound states between the sectors, such as a $\tilde{e}+W$? Or other spin dyons such as q-$\tilde{q}$ or $\tilde{e}$-$\mu$ bound states?

*19) CP-violation:* We suggest this Ansatz will contribute to the understanding of CP violation by the phases in the mixing of the neutrals and their same mass superpartners. The preponderance of matter over antimatter may be explained by the asymmetry of the handedness and the spin between the two sectors. Whether chromomagnetic matter affects the strong-CP problem, or monopoles affect axion-photon conversions is unknown. The duality between chromoelectric and chromomagnetic charges may obviate the need for an axion.

*20) Photinos as Dark Energy:* We introduce and sketch out an argument for a conjecture: Massless spinor $\tilde{\gamma}$, with frequencies approaching zero and wavelengths the size of the universe are natural candidates for Dark Energy, as a result of Fermi-pressure from a gas of relativistic massless spinor particles. If the density of $\tilde{\gamma}$, or pairs degenerate with $\gamma$, is similar to or exceeds that of $\gamma$, then this is plausible, as U/V~$10^{-28-29}$ g/cc by a crude over-estimate: assume the average photino energy is the same as the blackbody temperature. The average energy of a fully relativistic Fermion gas is $\langle E \rangle = \frac{3}{4} E_F$. $E_F = hc \left[ \frac{N}{V} \frac{3}{8\pi} \right]^{1/3}$, so n=N/V~$8 \times 10^9$ $\tilde{\gamma}$/m$^3$. The Fermi pressure $P_F$ = 3/4 $E_F$ n ~ $10^6$ eV/m$^3$. We emphasize that the fully relativistic Fermi pressure is dependent on the *number* density and not the energy density. For radiation, the universe energy density evolves as a$^{-4}$, and as a$^{-3}$ for matter, where a is the scale parameter; however, fermion $\tilde{\gamma}$ *number* density is more appropriate for scaling like matter, rather than bosonic radiation scaling, and possibly even as a$^{-2}$ or a$^{-1}$. If the photon is a linear combination or superposition of $\tilde{\gamma}$, similar to a linear-polarized photon as a superposition of two states, each with one photon with each circular polarizations, then in a toy model, if a photon has energy E, the photon wavefunction can be expressed as $\Psi(x,t) \sim e^{i/h\left(\frac{E-\delta}{c}x - (E-\delta)t\right)} + e^{i/h\left(\frac{E+\delta}{c}x - (E+\delta)t\right)}$, the sum of 2 $\tilde{\gamma}$ wavefunctions sharing a total energy E with the fluctuation $\delta$ in energy between them, with $\delta \to 0$, and $\delta/E <1$, similar to a linearly polarized photon is a superposition of two states, each with one photon with each circular polarizations. This is ok since $\tilde{\gamma}$ are massless and co-moving with $\gamma$. The fluctuation energy $\delta$ can be thought of as analogous to the arbitrarily small energy difference that creates bosonic cooper pairs, or the carrier of a force binds together 2 fermions. But as the photon redshifts or down-scatters, the wavelength increases, and the photinos can decohere, and the number of photinos is not constant. This should be observable in photon-photon scattering[20]. Therefore the number of photinos themselves is not conserved, rather, increasing; as their individual energy decreases with time, and photino states with longer wavelengths evolve – a single photino state can resolve into 3 photinos with lower energy, analogous to a heat pulse thermal diffusivity, creating more scalar or vector phonons or spin waves into larger volumes, but which would behave quite differently if fermionic. Additionally, the cosmos also is constantly making new photinos by: a) decreasing energy per photon/photino by scattering processes; b) $\gamma\gamma \leftrightarrow \tilde{\gamma}\tilde{\gamma}$ and $\gamma\gamma \leftrightarrow 4\tilde{\gamma}$ – oscillations between the mixed states of photons and photinos (depending on the exact description of the photino states), creating more photinos of lower energy; c) processes producing photinos, such as photon interactions with black holes, splitting a photon into an infalling photino and an escaping photino, and energetic processes creating $\gamma$ or $\nu$ or $\tilde{\nu}$ at sufficient intensity to create copious photinos (AGN, supernova). One may therefore consider photinos as constantly being "injected" over time to fill the expanding allowed fermion wavefunctions as the universe expands; unlike photons, these "new" photinos will try to occupy energy levels already occupied, which will produce a reaction force against the universe as the source of new photinos – negative pressure. As the size of the photino wavefunctions increases, and the numbers of photinos increase, the number which can overlap per unit volume may stay nearly the same ($a^0$) or decrease far less rapidly than radiation, say, $a^{-1}$ or $a^{-2}$. If this conjecture is true, it is in effect the negative pressure, which is arrived at by SUSY, the existence of degenerate SUSY states, as the number of allowed wavefunctions evolve to longer wavelengths to fit into the universe. This can be viewed of as a negative pressure creating a constant energy density.

*Discussion:*
Experimentally, electric charge is quantized, and the existence of any magnetic monopoles in nature provides a compelling reason. The t'Hooft-Polyakov insight is also a strong motivator for the existence of monopoles. But Nature would be especially perverse if only a single type of magnetic monopole existed; rather, in this model, magnetism is treated in one view identically with electricity, with a rich magnetic sparticle spectrum. Experimentally, (intrinsic) angular momentum and charge are quantized at the point interaction level, but energy-momentum are not fully quantized. It is therefore attractive to interchange charges and chirality when bosons and fermions are interchanged, rather than adjust masses. We have assumed that since color charge is quantized, that color magnetism is quantized, and part of the SUSY transform – but this may not be necessary. This mcSUSY model provides candidates for Dark Matter and Dark Energy, and yet provides a compellingly nearly invisible SUSY sector, while cancelling the divergences in the Higgs mass, and without R-parity and soft symmetry breaking. It also is a motivation to study non-perturbative effects in QED/QMD, QMD with scalars, and chromomagnetic gluino/squark systems. To whit, it is crucial whether a (pseudo)scalar $\tilde{e}$ with a magnetic charge at least as large as the minimal monopole charge, and with the mass of the electron, is already ruled out by precision experiments - tests of QED and the standard model. We assert that this is not clear-cut at present, and that it is possible that a monopole spectrum has remained hidden. We have not specified all the dimension-4 or greater couplings in this model, nor the size of the least magnetic monopole charges as multiples of $e/2\alpha$. How this ansatz with magnetism and wrong-handedness might be made to be imbedded in Grand Unification, quantum gravity, or in string or brane theories with more than 4 space-time dimensions, similar to attempts popular with extended supersymmetries, may be an interesting problem. We summarize some of the crucial issues: a) Identify existing data or phenomena that may already rule this SUSY out; b) Identify possible new experimental signatures on earth or in cosmology/astroparticle physics to test this proposal; c) In general, the consequences for or constraints from cosmology and the evolution of the universe – Dark Matter and Dark Energy; d) Formalize the allowed magnetic charges of the squarks relative to the sleptons; e) formalize the extended supersymmetry algebras and generators that would enable this form of supersymmetry; f) Whether there is a compelling theoretical demonstration, despite the breakdown of perturbation theory, that scalar magnetic charges are (quasi-) confined when pair-produced, and estimates of the production cross-sections, perhaps by borrowing from lattice QCD methods; g) Existence and properties of chromomagnetic/magnetic squark/gluino matter states, including cross sections of g=0 shadrons with ordinary matter; i.e, can chromomagnetic matter exist? Is it scalar like the squarks, or spinor like the gluino? Does magnetic energy prevent some squark bindings?; h) Modifications of neutrino oscillation phenomena, and whether sterile sneus already fit into experiments; i) Properties of non-perturbative magnetically bound states of the charged leptons; k) Existence of or properties of shydrogen; l) And many others not listed here but that are clearly inferable.  An aside: we have not specified whether the SUSY sector masses are gravi-magnetic (nor discussed the gravitino in this note); if gravimagnetic "poles" exist, perhaps that is why matter is also quantized, similar to Dirac's argument for magnetism, albeit different than the excellent arguments for quantization from compact groups.

*Acknowledgements:* We thank Fairfield University for support, and Les Schaffer, Grekim Jennings, Yasar Onel, and Jim Rich for helpful discussions and encouragement.